\RequirePackage{ifpdf}
\ifpdf 
\documentclass[pdftex]{sigma}
\else
\documentclass{sigma}
\fi

\begin{document}

\allowdisplaybreaks

\renewcommand{\thefootnote}{$\star$}

\renewcommand{\PaperNumber}{026}

\FirstPageHeading

\ShortArticleName{Spectral Distances: Results for Moyal Plane and Noncommutative Torus}

\ArticleName{Spectral Distances: Results for Moyal Plane\\ and Noncommutative Torus\footnote{This paper is a contribution to the Proceedings of
the XVIIIth International Colloquium on Integrable Systems and Quantum
Symmetries (June 18--20, 2009, Prague, Czech Republic).  The full
collection is
available at
\href{http://www.emis.de/journals/SIGMA/ISQS2009.html}{http://www.emis.de/journals/SIGMA/ISQS2009.html}}}

\Author{Eric CAGNACHE and Jean-Christophe WALLET}

\AuthorNameForHeading{E. Cagnache and J.-C. Wallet}

\Address{Laboratoire de Physique Th\'eorique, B\^at.\ 210, CNRS, Universit\'e Paris-Sud 11,\\
F-91405 Orsay Cedex, France}
\Email{\href{mailto:eric.cagnache@th.u-psud.fr}{eric.cagnache@th.u-psud.fr}, \href{mailto:jean-christophe.wallet@th.u-psud.fr}{jean-christophe.wallet@th.u-psud.fr}}

\ArticleDates{Received October 31, 2009, in f\/inal form March 20, 2010;  Published online March 24, 2010}

\Abstract{The spectral distance for noncommutative Moyal planes is considered in the framework of a non compact spectral triple recently proposed as a possible noncommutative analog of non compact Riemannian spin manifold.  An explicit formula for the distance between any two elements of a particular class of pure states can be determined. The corresponding result is discussed. The existence of some pure states at inf\/inite distance signals that the topology of the spectral distance on the space of states is not the weak $*$~topology. The case of the noncommutative torus is also considered and a formula for the spectral distance between some states is also obtained.}


\Classification{58B34; 46L52; 81T75}

\newcommand{\lrtimes}{\super{\ltimes}{\rtimes}}
\newcommand\caA{{\mathcal A}}
\newcommand\bbbone{\mathbb{I}}
\newcommand\caB{{\mathcal B}}
\newcommand\caG{{\mathcal G}}
\newcommand\dlde{ {\cal{D}}_{L^2} }
\newcommand\AD{{\mathcal A_{\mathcal D}}}
\newcommand\caM{{\mathcal M_\theta}}
\newcommand\caS{{\mathcal S}}
\newcommand\caU{{\mathcal U}}
\newcommand{\hd}{\hat{\mathrm{d}}}
\newcommand\caZ{{\mathcal Z}}
\newcommand\wx{{\widetilde x}}
\newcommand\gone{{ \mathchoice {1\mskip-4mu\mathrm{l} } {1\mskip-4mu\mathrm{l} }{1\mskip-4.5mu\mathrm{l} } {1\mskip-5mu\mathrm{l}} }}
\newcommand\gR{{\mathbb R}}
\def\gC{{\mathbb C}}
\def\ksu{{\mathfrak{su}}}
\def\gR{{\mathbb R}}
\def\ksl{{\mathfrak{sl}}}
\def\ksu{{\mathfrak{su}}}
\newcommand\gK{{\mathbb K}}
\newcommand\gN{{\mathbb N}}
\newcommand\gZ{{\mathbb Z}}
\newcommand\Omr{{\underline\Omega_\varepsilon}}
\newcommand\algzero{{\mathsf 0}}
\newcommand\algA{{\mathbf A}}
\newcommand\algrA{{\mathbf A^\bullet}}
\newcommand\algB{{\mathbf B}}
\newcommand\algrB{{\mathbf B^\bullet}}
\newcommand\modM{{\boldsymbol M}}
\newcommand\modrM{{\boldsymbol M^\bullet}}
\newcommand\kg{{\mathfrak g}}
\newcommand\hR{{F}}
\newcommand\DOM{\textup{Dom}}
\newcommand\SPEC{\textup{Sp}}
\newcommand\adrep{\textup{ad}}
\newcommand\der{{\text{\textup{Der}}}}
\newcommand\kX{{X}}
\newcommand\kY{{ Y}}
\newcommand\del{{\partial}}
\newcommand\delbar{{{\bar{\partial}}}}
\def\exter{{\textstyle\bigwedge}}
\newcommand\kS{{\mathfrak S}}
\newcommand\eps{{\varepsilon}}
\newcommand\ad{{\text{\textup{ad}}}}
\newcommand\nog{{\text{\textup{ng}}}}
\newcommand\gr{{\text{\textup{gr}}}}
\newcommand\inv{{\text{\textup{inv}}}}
\newcommand{\grast}{\bullet}
\newcommand\fois{\mathord{\cdot}}
\newcommand\Der{{\text{\textup{Der}}}}
\newcommand\Int{{\text{\textup{Int}}}}
\newcommand\out{{\text{\textup{Out}}}}
\newcommand\Out{{\text{\textup{Out}}}}
\newcommand\dd{{\text{\textup{d}}}}
\newcommand\invar{{\text{\textup{inv}}}}
\newcommand{\omi}[1]{\buildrel { \buildrel{#1}\over{\vee} } \over .}

\renewcommand{\thefootnote}{\arabic{footnote}}
\setcounter{footnote}{0}

\section{Introduction}

In noncommutative geometry (for reviews, see e.g.~\cite{CONNES, CM, LANDI, GRACIAVAR}), the notion of spectral triple gives rise naturally to the notion of spectral distance \cite{CONNES, CONNES1,CONNES2}. This spectral distance is def\/ined between two states $\varphi_1$ and $\varphi_2$ of the algebra ${\cal{A}}$ involved in the spectral triple\footnote{The algebra itself or its $C^*$-completion.}. Namely, one has
\begin{gather*}
d(\varphi_1,\varphi_2)\equiv\sup_{a\in{\cal{A}}}\big\{|(\varphi_1-\varphi_2)(a) |\ /\ ||[D,\pi(a)||_{\rm op}\le1 \big\},
\end{gather*}
where $D$ and $\pi(a)$ are respectively the Dirac operator and the representation of $a\in{\cal{A}}$ on the Hilbert space ${\cal{H}}$ that are other constituents of the spectral triple. When this latter encodes the usual commutative geometry of a f\/inite dimensional compact connected Riemann spin manifold, the spectral distance between two pure states (i.e.\ points) coincides with the geodesic distance.

The spectral distance between pure states can therefore be viewed as a natural noncommutative analog of the geodesic distance between points, while the actual interpretation of the spectral distance between two non pure states is not so clear at the present time. For a recent discussion on this point, see \cite{Andreamartinetti}. The properties of the spectral distance between two non-pure states have been examined closely by Rief\/fel in~\cite{Rieffel1} where it is indicated that the determination of the spectral distance between pure states is not suf\/f\/icient to exhaust the full metric information involved in the expression for $d(\varphi_1,\varphi_2)$ given above. The f\/irst works devoted to the explicit study of the spectral distance and its related properties, including some explicit determination of a distance formula focused on lattice geometries \cite{Muller1,Muller2,LIZZI}. This motivated further works inspired by theoretical physics and dealing with other simple noncommutative geometries \cite{Mart1,Martwulk,Mart2,Mart3}. These studies have def\/initely provided more insight on properties of spectral distances. However, they are related to a rather limited number of situations, namely almost commutative geometries, lattice geometries and f\/inite-dimensional noncommutative algebras, for which the technical aspects ruling the actual construction of the distance formula can be suitably mastered. Unfortunately, even for slightly more sophisticated noncommutative spaces, technical dif\/f\/iculties become rapidly overwhelming. This feature {\it a priori} prevents any explicit determination of distance formula to be obtained unless some simplifying reformulation is found. This basically explains why other examples of spectral distance formulas obtained within noncommutative spaces dif\/ferent from those mentioned above are lacking so far.

In Section~\ref{spectrdist} of this paper, the spectral distance for two-dimensional noncommutative Moyal plane is considered. The general mathematical properties of the Moyal spaces are well explained and detailed in e.g.~\cite{Gracia-Bondia:1987kw, Varilly:1988jk, GRACIAVAR}. In~\cite{GAYRAL2004}, a non compact spectral triple has been proposed as a reasonable noncommutative generalization of non compact Riemannian spin manifold (see also \cite{MARSE1,MARSE2}). The corresponding relevant material is collected in Subsection~\ref{moyal} to make the paper self-contained. Within the framework of the above triple, an explicit formula for the spectral distance between any two elements of a particular class of pure states can actually be computed, thanks to the use of simple algebraic properties of the set of elements $a\in{\cal{A}}$ verifying $||[D,\pi(a)]||_{\rm op}\le1$. The corresponding analysis is given in Subsection \ref{distmoy}. The analysis uses heavily the so called matrix base \cite{Gracia-Bondia:1987kw,Varilly:1988jk} which simplif\/ies the expressions involving the associative (non local) Moyal product, thanks to the existence of a Fr\'echet algebra isomorphism between the algebra of Schwartz functions of $\mathbb{R}^2$ and the algebra of rapidly decreasing sequences (see Proposition~\ref{matrix-base} below). In Section \ref{disc}, we discuss the results. The case of another non compact spectral triple proposed in~\cite{GAYRAL2004} is brief\/ly commented and expected to yields similar results. The spectral distance formula between any two elements of the particular class of pure states verif\/ies a triangular equality. There exists a family of pure states that are at inf\/inite distance from each other as well as at inf\/inite distance from some classes of pure states, including those for which the spectral distance formula constructed in Subsection~\ref{distmoy} is valid. Some related consequences are discussed. The existence of some pure states that are at inf\/inite distance implies that the topology induced by the spectral distance on the space of states of the algebra involved in the triple is not the weak $*$~topology, as discussed in~\cite{4-paper1}. It ref\/lects the fact that the spectral triple proposed in \cite{GAYRAL2004} is not a compact quantum metric space in the sense of~\cite{Rieffel5}. Recall that having weak $*$~topology as topology for the spectral distance on the space of states is a~condition to have compact quantum metric spaces as def\/ined in~\cite{Rieffel5}. Notice that a modif\/ied version of the considered non compact spectral triple that corresponds to a Rief\/fel compact quantum metric space has been proposed in~\cite{4-paper1}. Finally, we note that the technical lemma given in Subsection~\ref{distmoy} can be actually adapted to the algebra of the noncommutative torus. This gives rise immediately to the explicit construction of a spectral distance formula for the noncommutative torus between the tracial state and some vector states. The conclusions are given in Section~\ref{section4}.

\section{Spectral distance and Moyal non compact spin geometries}\label{spectrdist}

\subsection{General properties}\label{moyal}

In this subsection, we collect the essential material that will be used in the sequel. The main properties of the Moyal machinery can be found in e.g.~\cite{Gracia-Bondia:1987kw,Varilly:1988jk, GRACIAVAR} to which we refer for more details. An extension of the Connes spectral triple to the non compact case to which we will refer heavily thorough this paper has been proposed in~\cite{GAYRAL2004}. The corresponding action func\-tionals and spectral actions have been considered in \cite{MARSE1,MARSE2}. Constructions of various derivation based dif\/ferential calculi on Moyal algebras have been carried out in \cite{WAL1,WAL2, WAL3}, together with applications to the construction of gauge-Higgs models on Moyal spaces. In this work, we will only consider the two-dimensional case.

Let ${\cal{S}}({\mathbb{R}}^2)\equiv{\cal{S}}$ be the Fr\'echet space of complex-valued Schwartz functions on ${\mathbb{R}}^2$ and ${\cal{S}}^\prime({\mathbb{R}}^2)\equiv{\cal{S}}^\prime$ its topological dual space. $L^2({\mathbb{R}}^2)$ denotes as usual the Hilbert space of square integrable functions on ${\mathbb{R}}^2$. $||\cdot||_2$ is the $L^2({\mathbb{R}}^2)$ norm.
\begin{proposition}[see e.g.~\cite{Gracia-Bondia:1987kw, Varilly:1988jk}]
The associative bilinear Moyal $\star$-product is def\/ined for any $a,b\in{\cal{S}}$ by:
\begin{gather}
\star: \ {\cal{S}}\times{\cal{S}}\to{\cal{S}},\nonumber\\
(a\star b)(x)=\frac{1}{(\pi\theta)^2}\int d^2yd^2z\, a(x+y)b(x+z)e^{-i2y\Theta^{-1}z},\nonumber\\
 y\Theta^{-1}z\equiv y^\mu \Theta^{-1}_{\mu\nu}z^\nu,\qquad  \Theta_{\mu\nu}=\theta\begin{pmatrix} 0&1 \\ -1& 0 \end{pmatrix},\label{eq:moyal}
\end{gather}
with $\theta\in\mathbb{R}$, $\theta\ne0$. The complex conjugation is an involution for the $\star$-product and one has the faithful trace given by $\int d^2x\ (a\star b)(x)=\int d^2x\ (b\star a)(x)=\int d^2x \, a(x)b(x)$ and the Leibniz rule $\partial_\mu(a\star b)=\partial_\mu a\star b+a\star\partial_\mu b$, $\forall\, a,b\in{\cal{S}}$. ${\cal{A}}\equiv({\cal{S}},\star)$ is a non unital involutive Fr\'echet algebra.
\end{proposition}

A part of our analysis will use the matrix base whose relevant properties are summarized by the following proposition.
\begin{proposition}[see e.g.~\cite{Gracia-Bondia:1987kw, Varilly:1988jk}]\label{matrix-base}
We set $X^{\star n}\equiv X\star X\star\cdots \star X$ where $n$ factors are involved and define $[f,g]_\star\equiv f\star g-g\star f$. The matrix base is the family of functions $\{f_{mn}\}_{m,n\in\mathbb{N}}\subset{\cal{S}}\subset L^2({\mathbb{R}}^2)$ defined by the Wigner eigenfunctions of the two-dimensional harmonic oscillator. Set ${\bar{z}}={{1}\over{{\sqrt{2}}}}(x_1-ix_2)$, $z={{1}\over{{\sqrt{2}}}}(x_1+ix_2)$ and denote by $\langle\cdot,\cdot\rangle$ the inner product on $L^2({\mathbb{R}}^2)$.

$i)$ One has the relations:
\begin{gather}
f_{mn}={{1}\over{(\theta^{m+n}m!n!)^{1/2}}}{\bar{z}}^{\star m}\star f_{00}\star z^{\star n},\qquad  f_{00}=2e^{-2H/\theta},\qquad H=\tfrac{1}{2} \big(x_1^2+x_2^2\big),\nonumber
\\
f_{mn}\star f_{pq}=\delta_{np}f_{mq},\qquad f_{mn}^*=f_{nm},\qquad  \langle f_{mn},f_{kl} \rangle=(2\pi\theta)\delta_{mk}\delta_{nl}.\label{algmatrixbase}
\end{gather}

$ii)$ There is a Fr\'echet algebra isomorphism between ${\cal{A}}\equiv({\cal{S}},\star)$ and the matrix algebra of decreasing sequences $(a_{mn})$, $\forall\, m,n\in{\mathbb{N}}$ defined by $a=\sum_{m,n}a_{mn}f_{mn}$, $\forall\, a\in{\cal{S}}$, such that the semi-norms $\rho_k^2(a)\equiv\sum_{m,n}\theta^{2k}(m+{{1}\over{2}})^k(n+{{1}\over{2}})^k|a_{mn}|^2$, $\forall \, k\in{\mathbb{N}}$, for the topology of the latter algebra are finite.
\end{proposition}

The $\star$ product \eqref{eq:moyal} can be extended to spaces larger than ${\cal{S}}$, using duality and continuity of the $\star$-product on~${\cal{S}}$. A convenient starting point that we recall for the sake of completeness is to introduce the family of spaces ${\cal{G}}_{s,t}$, $s,t\in{\mathbb{R}}$, ${\cal{S}}\subset{\cal{G}}_{s,t}\subset{\cal{S}}^\prime$ densely and continuously, def\/ined~as
\begin{gather*}
{\cal{G}}_{s,t}=\Bigg\{ a=\sum_{m,n\in{\mathbb{N}}}a_{mn}f_{mn}\in{\cal{S}}^\prime\ / \ ||a ||_{s,t}^2=\sum_{m,n}\theta^{s+t}\big(m+\tfrac{1}{2}\big)^s\big(n+\tfrac{1}{2})^t|a_{mn}|^2<\infty\Bigg\} ,
\end{gather*}
with $a_{mn}={{1}\over{2\pi\theta}}\int d^2xa(x)f_{nm}(x)={{1}\over{2\pi\theta}}\int d^2x(a\star f_{nm})(x)$ where the $\star$-product has been extended to ${\cal{S}}^\prime\times{\cal{S}}\to{\cal{S}}^\prime$ by duality. Then, for any $a\in{\cal{G}}_{s,t}$ and any $b\in{\cal{G}}_{q,r}$, $b=\sum_{m,n}b_{mn}f_{mn}$, with $t+q\ge0$, the sequences $c_{mn}=\sum_{p}a_{mp}b_{pn}$, $\forall\, m,n\in{\mathbb{N}}$ def\/ine the functions $c=\sum_{m,n}c_{mn}f_{mn}$, $c\in{\cal{G}}_{s,r}$, as $||a\star b||_{s,r}\le||a||_{s,t}||b||_{q,r}$, $t+q\ge0$ and $|| a||_{u,v}\le ||a ||_{s,t}$ if $u\le s$ and $v\le t$. For more details, see e.g.~\cite{Gracia-Bondia:1987kw, Varilly:1988jk}.

\begin{remark}
Recall by the way that any radial function $a$ is diagonal in the matrix base, which means that it can be expanded as $a=\sum_{m\in\mathbb{N}}a_mf_{mm}$ where the diagonal Schwartz functions of the matrix base can be explicitly written as $f_{mm}=(-1)^m2L_m\big({{2r^2}\over{\theta}}\big)\exp\big({-}{{r^2}\over{\theta}}\big)$, $\forall\, m\in\mathbb{N}$, with $r^2=x_1^2+x_2^2$, ($L_m(x)$ is the Laguerre polynomial of order~$m$) \cite{Gracia-Bondia:1987kw, Varilly:1988jk}.
\end{remark}

Let $\dlde$ be the space of square integrable smooth functions $a\in L^2({\mathbb{R}}^2)\cap C^\infty({\mathbb{R}}^2)$ ha\-ving all their derivatives in $L^2({\mathbb{R}}^2)$. $(\dlde,\star)$ is a Fr\'echet $*$-algebra for the norms $q_m(a)=\max\limits_{|\alpha|<m}||\del^\alpha a ||_2$, $m\in\mathbb{N}$, for any $a\in\dlde$ \cite{LaurentSchw}. Let ${\cal{B}}$ be the space of functions $a\in C^\infty({\mathbb{R}}^2)$ that are bounded together with all their derivatives \cite{LaurentSchw}. Let $\algA_\theta=\{a\in{\cal{S}}^\prime\  /\ a\star b\in L^2({\mathbb{R}}^2)$, $ \forall \, b\in L^2({\mathbb{R}}^2)\}$. We set from now on ${\cal{A}}\equiv({\cal{S}},\star)$ and  ${\cal{A}}_1\equiv({\cal{B}},\star)$.

\begin{proposition}[\cite{Varilly:1988jk, GAYRAL2004, LaurentSchw}] \label{algebras} The following properties hold:

$i)$ $(\algA_\theta,\star)$ is a unital $C^*$-algebra with the operator norm $||\cdot||_{\rm op}$,
$||a||_{\rm op}=\sup\limits_{0\ne b\in L^2({\mathbb{R}}^2)}\big\{{{||a\star b||_2}\over{||b||_2}} \big\}$ for any $a\in\algA_\theta$, isomorphic to ${\cal{L}}(L^2({\mathbb{R}}^2))$ {\rm \cite{GAYRAL2004}}.

$ii)$ ${\cal{A}}_1$ is a Fr\'echet algebra for the semi-norms $p_m(a)=\max\limits_{| \alpha|<m}|| \partial^\alpha a||_\infty$, $m\in\mathbb{N}$. ${\cal{A}}_1$ is a~pre $C^*$-algebra and one has the inclusion ${\cal{A}}\subset(\dlde,\star)\subset{\cal{A}}_1\subset\algA_\theta$  $($Calder\'on--Vaillancourt theorems$)$.
\end{proposition}

We consider a non compact spectral triple of the type proposed in \cite{GAYRAL2004} as a reasonable noncommutative generalization of non compact Riemannian spin manifold. The corresponding axioms are discussed in \cite{GAYRAL2004} to which we refer for more details. The building data of this non compact spectral triple are
\begin{equation}
({\cal{A}},{\cal{A}}_1,{\cal{H}},D;J,\chi), \label{nctriple}
\end{equation}
in which the antiunitary operator $J$ and involution $\chi$ will not be relevant here. ${\cal{A}}_1\supset{\cal{A}}$ is a~preferred unitization of ${\cal{A}}$ in the sense of \cite{GAYRAL2004}. The main properties of \eqref{nctriple} that will be relevant for the computation of the spectral distance formula are collected below.

The Hilbert space ${\cal{H}}$ is ${\cal{H}}=L^2(\mathbb{R}^2)\otimes \mathbb{C}^2$. It is the Hilbert space of integrable square sections of the trivial spinor bundle ${\mathbb{S}}={\mathbb{R}}^2\otimes{\mathbb{C}}^2$ with Hilbert product
$\langle \psi,\phi\rangle=\int  d^2x(\psi_1^*\phi_1+\psi_2^*\phi_2)$ $\forall\, \psi,\phi\in{\cal{H}}$ with $\psi=(\psi_1,\psi_2)$, $\phi=(\phi_1,\phi_2)$.

We def\/ine now $\partial={{1}\over{{\sqrt{2}}}}(\partial_1-i\partial_2)$, ${\bar{\partial}}={{1}\over{{\sqrt{2}}}}(\partial_1+i\partial_2)$. In \eqref{nctriple}, $D$ is the unbounded Euclidean self-adjoint Dirac operator $D=-i\sigma^\mu\del_\mu$, densely def\/ined on $\DOM(D)=(\dlde\otimes{\mathbb{C}}^2)\subset{\cal{H}}$, where the $\sigma^\mu$'s verify $\sigma^\mu\sigma^\nu+\sigma^\nu\sigma^\mu=2\delta^{\mu\nu}$, $\forall\, \mu,\nu=1,2$ and span an irreducible representation of ${\mathbb{C}}l({\mathbb{R}}^2,\delta)$ ($\delta$ is the Euclidean metric). One can write
\begin{gather*}
\sigma^1=\begin{pmatrix} 0&1 \\ 1& 0 \end{pmatrix},\qquad \sigma^2=\begin{pmatrix} 0&i \\ -i& 0 \end{pmatrix},\qquad D=-i{\sqrt{2}}\begin{pmatrix} 0&\delbar \\ \del& 0 \end{pmatrix}.
\end{gather*}
The algebra ${\cal{A}}$ can be represented faithfully on ${\cal{B}}({\cal{H}})$, the space of bounded operators on ${\cal{H}}$. The representation $\pi:{\cal{A}}\to{\cal{B}}({\cal{H}})$ is def\/ined by
\begin{gather*}
\pi(a)=L(a)\otimes\bbbone_2,\qquad \pi(a)\psi=(a\star\psi_1,a\star\psi_2),\qquad \forall\, \psi=(\psi_1,\psi_2)\in{\cal{H}},\qquad \forall\, a\in{\cal{A}}.
\end{gather*}
Here, $L(a)$ is the left multiplication operator by any $a\in{\cal{A}}$ and $\pi(a)$ and $[D,\pi(a)]$ are bounded operators on ${\cal{H}}$ for any $a\in{\cal{A}}$. $\pi(a)\in{\cal{B}}({\cal{H}})$ stems from $||a\star b ||_2\le ||a ||_2 || b||_2 $ for any $a,b\in L^2({\mathbb{R}}^2)$.

$[D,\pi(a)]\subset{\cal{B}}({\cal{H}})$ follows from $||[D,\pi(a)]||_{\rm op}={\sqrt{2}}\max(||L(\del a)||_{\rm op},||L(\delbar a)||_{\rm op})$ (see equation~(\ref{normop}) below) and ${\cal{A}}\subset L^2(\mathbb{R}^2)$. Note that in the present version of non compact spectral triple, $\pi(a)(D-\lambda)^{-1}$ is compact for any $\lambda\notin {\rm Sp}\,(D)$ \cite{GAYRAL2004} while for a canonical compact spectral triple, $D$ has compact resolvent.

\begin{remark}
Notice that, in view of the inclusion property given in $ii)$ of Proposition \ref{algebras}, one has $\pi(a)\in{\cal{B}}({\cal{H}})$ and $[D,\pi(a)]\in{\cal{B}}({\cal{H}})$ for any $a\in{\cal{A}}_1$.
\end{remark}

\subsection{Spectral distance on the Moyal plane}\label{distmoy}

It is known that the notion of spectral distance is related naturally to the notion of spectral triple. For general reviews on noncommutative spectral geometry and related notion of spectral distance, see e.g.~\cite{CONNES,LANDI}. So far, relatively few works devoted to the explicit study of the spectral distance and including explicit determination of spectral distance formula have appeared in the literature. Among these works, the f\/irst one focused on lattice geometries \cite{Muller1,Muller2,LIZZI}, followed by studies inspired by theoretical physics and dealing with other simple noncommutative geometries~\cite{Mart1,Martwulk,Mart2,Mart3}.

\begin{definition}\label{spectraldist}
The spectral distance between any two states $\omega_1$ and $\omega_2$ of ${\bar{ {\cal{A}}}}$ is def\/ined by
\begin{equation}
d(\omega_1,\omega_2):=\sup_{a\in{\cal{A}}}\big\{|\omega_1(a)-\omega_2(a)|;\ ||[D,\pi(a)] ||_{\rm op}\le1\big\},\label{spectralformule}
\end{equation}
where $||\cdot||_{\rm op}$ is the operator norm for the representation of ${\cal{A}}$ in ${\cal{B}}({\cal{H}})$.
\end{definition}

For convenience, we set the following notation:
\begin{definition}\label{unit-ball}
We def\/ine ${\cal{B}}_1:=\big\{a\in{\cal{A}}\ /\ ||[D,\pi(a)] ||_{\rm op}\le1  \big\}$.
\end{definition}

\begin{remark} One can check that \eqref{spectralformule} obeys the def\/ining mathematical properties for a distance although inf\/inite values are now allowed for $d(\omega_1,\omega_2)$. The spectral distance between pure states can be viewed as a natural noncommutative analog of the geodesic distance between two points. Indeed, recall that the spectral distance for the (commutative) canonical spectral triple reproducing the geometry of f\/inite dimensional compact connected Riemann spin manifold coincides with the related geodesic distance \cite{CONNES, LANDI}.  Note that \eqref{spectralformule} extends the notion of distance to non-pure states, i.e.\ objects that are not analog to points. The actual ``interpretation'' of the spectral distance between two non pure states is not quite clear at the present time. Mathematical investigation reported in~\cite{Rieffel1} has indicated that the determination of the spectral distance between pure states is not enough to exhaust the full metric information involved in~\eqref{spectralformule}. Notice that a relationship with the Wasserstein distance of order~1 between probability distributions on a metric space and the spectral distance has been exhibited recently in~\cite{Andreamartinetti}.
\end{remark}

So far, the only few available explicit computations of spectral distance formula have been performed within ``relatively simple'' noncommutative geometries, namely almost commutative geometries, lattice geometries and f\/inite-dimensional noncommutative algebras, in which the main technical aspects involved in the construction of the distance formula can be dealt with. However, the extension to even slightly more sophisticated noncommutative spaces was soon realized to be out of reach in the absence of suitable simplifying reformulation that would allow the increasing technical dif\/f\/iculties to be still mastered. It turns out that such a simplif\/ication can be performed within the Moyal plane by using the matrix base, $i)$ of Proposition \ref{matrix-base}, thanks to the Fr\'echet algebra isomorphism it def\/ines between ${\cal{A}}$ and the matrix algebra of rapidly decreasing sequences as given in $ii)$ of Proposition \ref{matrix-base}. This permits one to actually compute a~spectral distance formula between any two elements of a particular class of pure states.

The $C^*$-completion of ${\cal{A}}$ relative to the operator norm is denoted by ${\bar{ {\cal{A}}}}$. The determination of the pure states of ${\bar{ {\cal{A}}}}$ can be done conveniently by exploiting some properties of the matrix base. The starting point is the observation is that any of the vector states def\/ined by any element~$f_{mn}$ of the matrix base of Proposition~\ref{matrix-base} depends only on the f\/irst index $m\in{\mathbb{N}}$, thanks to the relations~\eqref{algmatrixbase} and that ${\cal{A}}$ can be actually interpreted as an algebra of compact operators acting on some Hilbert space. Then, the following property follows:

\begin{proposition}\label{purestates}
The pure states of ${\bar{ {\cal{A}}}}$ are the vector states $\omega_\psi:{\bar{ {\cal{A}}}}\to\mathbb{C}$ defined by any unit vector $\psi\in L^2(\mathbb{R}^2)$ of the form $\psi=\sum_{m\in\mathbb{N}}\psi_mf_{m0}$, $\sum_{m\in\mathbb{N}}| \psi_m|^2={{1}\over{2\pi\theta}}$ and one has
\begin{gather*}
\omega_\psi (a) \equiv\big\langle(\psi,0),\pi(a)(\psi,0)\big\rangle=2\pi\theta\sum_{m,n\in{\mathbb{N}}} \psi^*_m\psi_n  a_{mn} .
\end{gather*}
\end{proposition}

\begin{proof}
Let ${\cal{H}}_0$ be the Hilbert space spanned by the family $(f_{m0})_{m\in{\mathbb{N}}}$. For any $a=\sum_{m,n}a_{mn}f_{mn}$ $\in\, {\cal{A}}$, one has $\sum_p||L(a)f_{p0} ||_2^2=\sum_{p,m}|a_{pm} |^2=|| a||_2^2<\infty$. Then, $L(a)$ is a Hilbert--Schmidt operator on ${\cal{H}}_0$ and therefore is compact on ${\cal{H}}_0$. Let $\pi_0$ be this representation of ${\cal{A}}$ on ${\cal{H}}_0$ and $\overline{\pi_0({\cal{A}})}$ be the completion of $\pi_0({\cal{A}})$. One has ${\overline{\pi_0({\cal{A}})}}\subseteq {\cal{K}}({\cal{H}}_0)$. $\pi_0({\cal{A}})$ involves all f\/inite rank operators. Then ${\overline{\pi_0({\cal{A}})}}\supseteq {\cal{K}}({\cal{H}}_0)$ and so ${\overline{\pi_0({\cal{A}})}}={\cal{K}}({\cal{H}}_0)$. This latter has a unique irreducible representation (up to unitary equivalence) and the corresponding pure states are exactly given by vectors states def\/ined by any unit vector $\psi=\sum_{m\in\mathbb{N}}\psi_mf_{m0}\in{\cal{H}}_0$.
\end{proof}

To study the spectral distance as given in Def\/inition~\ref{spectraldist}, we need to characterize conveniently the set ${\cal{B}}_1$, Def\/inition~\ref{unit-ball}. This can be achieved by exploiting algebraic relations among the coef\/f\/icients of the expansion in the matrix base of any $a$ and $\del a$ and $\delbar a$. The relevant features are summarized in the following proposition.

\begin{proposition}\label{relationalpha} \quad

$i)$ The $f_{mn}$'s and their derivatives satisfy:
\begin{gather}
\partial f_{mn}={\sqrt{{{n}\over{\theta}}}}f_{m,n-1}-{\sqrt{{{m+1}\over{\theta}}}}f_{m+1,n},\nonumber\\
 {\bar{\partial}} f_{mn}={\sqrt{{{m}\over{\theta}}}}f_{m-1,n}-{\sqrt{{{n+1}\over{\theta}}}}f_{m,n+1}, \qquad \forall\, m,n\in{\mathbb{N}}.\label{derivfmn}
\end{gather}

$ii)$ For any $a\in{\cal{A}}$, $a=\sum_{m,n}a_{mn}f_{mn}$, we define $\del a\equiv\sum_{m,n}\alpha_{mn}f_{mn}$ and $\delbar a\equiv\!\sum_{m,n}{{\beta}}_{mn}f_{mn}$.

\quad \ $a)$ The following relations hold:
\begin{gather}
\alpha_{m+1,n}={\sqrt{{{n+1}\over{\theta}}}}a_{m+1,n+1}-{\sqrt{{{m+1}\over{\theta}}}}a_{m,n},\nonumber\\ \alpha_{0,n}={\sqrt{{{n+1}\over{\theta}}}}a_{0,n+1},\qquad  \forall\, m,n\in{\mathbb{N}} ,\label{alphamn}
\\
{{\beta}}_{m,n+1}={\sqrt{{{m+1}\over{\theta}}}}a_{m+1,n+1}-{\sqrt{{{n+1}\over{\theta}}}}a_{m,n},\nonumber\\ {{\beta}}_{m,0}={\sqrt{{{m+1}\over{\theta}}}}a_{m+1,0},\qquad \forall\, m,n\in{\mathbb{N}}. \label{betamn}
\end{gather}

\quad \ $b)$ One has the inversion formula:
\begin{equation}
a_{p,q}=\delta_{p,q}a_{0,0}+\sqrt{\theta}\,\sum_{k=0}^{\min(p,q)}
\frac{\alpha_{p-k,q-k-1}+\beta_{p-k-1,q-k}}{\sqrt{p-k}+\sqrt{q-k}}, \qquad \forall \, p,q\in{\mathbb{N}},\qquad p+q>0 . \label{poincarelemma}
\end{equation}
\end{proposition}

\begin{proof}
The proof of the property $i)$ is a standard calculation. The property $iia)$ follows from~\eqref{derivfmn} of $i)$ and a simple calculation. To prove $iib)$, one combines \eqref{alphamn} and \eqref{betamn} to obtain
\begin{gather*}
a_{p+1,q+1}=
a_{p,q}+\sqrt{\theta}\,\frac{\alpha_{p+1,q}+\beta_{p,q+1}}{\sqrt{p+1}+\sqrt{q+1}}.
\end{gather*}
This yields by induction
\begin{gather*}
a_{p,q}=
\begin{cases}
\displaystyle a_{0,q-p}+\sqrt{\theta} \sum_{k=0}^{p-1}
 \frac{\alpha_{p-k,q-k-1}+\beta_{p-k-1,q-k}}{\sqrt{p-k}+\sqrt{q-k}} &\text{if}\ \ 0\leq p<q ,\\
\displaystyle a_{0,0}+\sqrt{\theta} \sum_{k=0}^{p-1}
\frac{\alpha_{p-k,p-k-1}+\beta_{p-k-1,p-k}}{2\sqrt{p-k}} &\text{if} \ \ 0\leq p=q ,\\
\displaystyle a_{p-q,0}+\sqrt{\theta}\,\sum_{k=0}^{q-1}
\frac{\alpha_{p-k,q-k-1}+\beta_{p-k-1,q-k}}{\sqrt{p-k}+\sqrt{q-k}} &\text{if}\ \ 0\leq q<p .
\end{cases}
\end{gather*}
Then, a further use of the second relations for $\alpha_{0,n}$ and $\beta_{0,n}$ in \eqref{alphamn} and \eqref{betamn} gives rise to $iib)$.
\end{proof}

The condition def\/ining ${\cal{B}}_1$, Def\/inition \ref{unit-ball} can be translated into constraints on the coef\/f\/icients the expansion of $\del a$ and $\delbar a$ in the matrix base. Indeed, observe that for any $\psi\in{\cal{H}}=L^2(\mathbb{R}^2)\oplus L^2(\mathbb{R}^2)$, $\psi=(\varphi_1,\varphi_2)$, one has
\begin{gather*}
[D,\pi(a)]\psi=-i(L(\partial_\mu a)\otimes\sigma^\mu)\psi=
-i{\sqrt{2}}\begin{pmatrix} \delbar a\star\varphi_2 \\ \del a\star\varphi_1 \end{pmatrix}=-i{\sqrt{2}}\begin{pmatrix} L(\del a)&0 \\ 0& L(\delbar a) \end{pmatrix}\begin{pmatrix} \varphi_2 \\ \varphi_1 \end{pmatrix}. 
\end{gather*}
This implies
\begin{equation}
 ||[D,\pi(a)]||_{\rm op}={\sqrt{2}}\max(||L(\del a)||_{\rm op},||L(\delbar a)||_{\rm op})={\sqrt{2}}\max(||\del a||_{\rm op},||\delbar a||_{\rm op})\label{normop}
\end{equation}
and therefore
\begin{gather*}
 ||[D,\pi(a)] ||_{\rm op}\le1\ \iff\ ||\del a||_{\rm op}\le{{1}\over{{\sqrt{2}}}}\qquad  \text{and} \qquad ||\delbar a||_{\rm op}\le{{1}\over{{\sqrt{2}}}}.
\end{gather*}
The following def\/inition f\/ixes the notation that will be used in this section.

\begin{definition}\label{purestatedef}
We denote by $\omega_m$ the pure state generated by the unit vector ${{1}\over{{\sqrt{2\pi\theta}}}}f_{m0}$, $\forall \, m\in{\mathbb{N}}$. For any $a=\sum_{m,n}a_{mn}f_{mn}\in{\cal{A}}$, one has $\omega_{m}(a)=a_{mm}$.
\end{definition}

We now prove a lemma that will permit us to actually determine the distance between any pure states $\omega_m$ introduced in Def\/inition \ref{purestatedef}.
\begin{lemma}\label{lemme2}
We set $\del a=\sum_{m,n}\alpha_{mn}f_{mn}$ and $\delbar a=\sum_{m,n}\beta_{mn}f_{mn}$, for any $a\in{\cal{A}}$ and any unit vector $\varphi=\sum_{m,n}\varphi_{mn}f_{mn}\in L^2(\mathbb{R}^2)$.

$i)$ Assume that $||[D,\pi(a)] ||_{\rm op}\le1$. Then, the following property $({\mathfrak{P}})$ holds:
\begin{gather*}
({\mathfrak{P}})\ \sum_{p}|\alpha_{mp}||\varphi_{pn}| \le{{1}\over{2{\sqrt{\pi\theta}}}}\qquad {and} \\ \sum_{p}|\beta_{mp}||\varphi_{pn}| \le{{1}\over{2{\sqrt{\pi\theta}}}},\qquad \forall\, \varphi\in{\cal{H}}_0,\qquad || \varphi||_2=1,\qquad \forall \, m,n\in{\mathbb{N}}.
\end{gather*}

$ii)$ If $({\mathfrak{P}})$ holds, then $|\alpha_{mn}|\le{{1}\over{{\sqrt{2}}}}$ and $|\beta_{mn}|\le{{1}\over{{\sqrt{2}}}}$, $\forall\, m,n\in{\mathbb{N}}$.

$iii)$ For any radial function $a\in{\cal{A}}$ $($i.e.\ $a_{mn}=0$ if $m\neq n)$, $||[D,\pi(a)] ||_{\rm op}\le1$ is equivalent to $|\alpha_{mn}|\le{{1}\over{{\sqrt{2}}}}$ and $|\beta_{mn}|\le{{1}\over{{\sqrt{2}}}}$, $\forall\, m,n\in{\mathbb{N}}$.

$iv)$ Let ${\hat{a}}(m_0):=\sum_{p,q\in\mathbb{N}}{\hat{a}}_{pq}(m_0)f_{pq}$, where ${\hat{a}}_{pq}(m_0)=\delta_{pq}{\sqrt{{{\theta}\over{2}}}}\sum_{k=p}^{m_0}{{1}\over{{\sqrt{k+1}}}}$ with fixed $m_0\in{\mathbb{N}}$. Let ${\cal{A}}_+$ denotes the set of positive elements of ${\cal{A}}$. Then, ${\hat{a}}(m_0)\in{\cal{A}}_+$ and $||[D,\pi({\hat{a}}(m_0))] ||_{\rm op}=1$ for any $m_0\in\mathbb{N}$.
\end{lemma}

\begin{proof}
If $||[D,\pi(a)] ||_{\rm op}\le1$, then $||\del a||_{\rm op}\le{{1}\over{{\sqrt{2}}}}$ and $||\delbar a||_{\rm op}\le{{1}\over{{\sqrt{2}}}}$. Using the matrix base, a~standard calculation yields, for any $\varphi\in{\cal{H}}_0$, $||\del a\star\varphi ||_2^2=2\pi\theta\sum_{m,n}|\sum_p\alpha_{mp}\varphi_{pn}|^2$. Owing to the def\/inition of $||\del a||_{\rm op}$, one obtains immediately $\sum_{m,n}|\sum_p\alpha_{mp}\varphi_{pn}|^2\le{{1}\over{4\pi\theta}}$ for any $\varphi\in{\cal{H}}_0$ with $\sum_{m,n}|\varphi_{mn}|^2={{1}\over{2\pi\theta}}$. This implies
\begin{equation}
\Big|\sum_p\alpha_{mp}\varphi_{pn}\Big|\le{{1}\over{2{\sqrt{\pi\theta}}}},\qquad \forall\,\varphi\in{\cal{H}}_0,\qquad || \varphi||_2=1,\qquad \forall \, m,n\in{\mathbb{N}} \label{PP1}
\end{equation}
together with a similar relation stemming from $||\delbar a||_{\rm op}$ with the $\alpha_{mn}$'s replaced by $\beta_{mn}$.

Now, $|\sum_p\alpha_{mp}\varphi_{pn}|\le{{1}\over{2{\sqrt{\pi\theta}}}}$ holds true for any $\varphi\in{\cal{H}}_0$ with $||\varphi||_2=1$ and one can construct~${\tilde{\varphi}}$ with $||{\tilde{\varphi}}||_2=|| \varphi||_2$ via $\alpha_{mp}{\tilde{\varphi}_{pn}}=|\alpha_{mp}||\varphi_{pn}|$. Then, this implies
\begin{equation}
\sum_{p}|\alpha_{mp}||\varphi_{pn}|\le{{1}\over{2{\sqrt{\pi\theta}}}},\qquad \forall\,\varphi\in{\cal{H}}_0,\qquad || \varphi||_2=1,\qquad \forall \, m,n\in{\mathbb{N}}. \label{PP2}
\end{equation}
Notice that \eqref{PP2} implies \eqref{PP1}. Similar considerations apply for the $\beta_{mn}$'s. This terminates the proof for property $i)$. The property $ii)$ is then a direct consequence of the property ${\mathfrak{P}}$.

To prove $iii)$, one has just to prove that any radial function $a$ such that $|\alpha_{mn}|\le{{1}\over{{\sqrt{2}}}}$ and $|\beta_{mn}|\le{{1}\over{{\sqrt{2}}}}$, $\forall \, m,n\in{\mathbb{N}}$ is in ${\cal{B}}_1$. One f\/irst observe that if $a$ is radial, one has $\alpha_{mn}=0$ if $m\neq n+1$ thanks to~\eqref{alphamn}. Then, for any unit vector $\psi\in{\cal{H}}_0$, one has
\begin{equation}
||\del{{a}}\star\psi||^2_2=2\pi\theta\sum_{p,q}\Big|\sum_r{{\alpha}}_{pr}\psi_{rq} \Big|^2=2\pi\theta\sum_{p,q}|{{\alpha}}_{p,p-1}\psi_{p-1,q} |^2\le\pi\theta\sum_{p,q\in{\mathbb{N}}}|\psi_{pq}|^2\label{smallnorm}
\end{equation}
so that $||\del{{a}}||_{\rm op}^2\le{{1}\over{2}}$ showing that ${{a}}$ is in ${\cal{B}}_1$. Similar considerations apply for $\beta_{mn}$. The pro\-per\-ty~$iii)$ is shown.

To prove that ${\hat{a}}(m_0)\in{\cal{A}}$ def\/ines a positive operator of ${\cal{B}}({\cal{H}})$ for any f\/ixed $m_0\in\mathbb{N}$, one has to show the following inequality $\langle \psi,\pi({\hat{a}}(m_0))\psi\rangle \ge 0$, $\forall\, \psi\in{\cal{H}}$, for any f\/ixed $m_0\in\mathbb{N}$. Set $\psi=(\varphi_1,\varphi_2)$, $\varphi_i\in L^2(\mathbb{R}^2)$, $i=1,2$ and $\varphi_i=\sum_{m,n\in\mathbb{N}}\varphi^i_{mn}f_{mn}$. A simple calculation yields
\begin{gather*}
\langle \varphi_i,{\hat{a}}(m_0)\star\varphi_i\rangle =2\pi\theta{\sqrt{{{\theta}\over{2}}}}\sum_{m\le m_0,n}|\varphi^i_{mn}|^2\left(\sum_{k=m}^{m_0}{{1}\over{{\sqrt{k+1}}}}\right),
\end{gather*}
and therefore, $\langle \varphi_i,{\hat{a}}(m_0)\star\varphi_i\rangle \ge0$, implying $\langle \psi,\pi({\hat{a}}(m_0))\psi\rangle =\sum_{i=1}^2\langle \varphi_i,{\hat{a}}(m_0)\star\varphi_i\rangle \ge0$, so that ${\hat{a}}(m_0)\in{\cal{A}}_+$ for any f\/ixed $m_0\in\mathbb{N}$.

Finally, notice that any positive element $a\in{\cal{A}}_+$ verif\/ies $a^\dag=a$ so that $(\partial a)^\dag=\delbar a$. Then, one obtains
from~\eqref{normop}: $||[D,\pi(a)] ||_{\rm op}={\sqrt{2}}||\del a ||_{\rm op}$. Now, by using $iia)$ of Proposition~\ref{relationalpha}, a standard calculation shows that the only non-vanishing coef\/f\/icients ${\hat{\alpha}}_{pq}$ in the matrix base expansion of $\del {\hat{a}}(m_0)$ satisfy ${\hat{\alpha}}_{p+1,p}=-{{1}\over{{\sqrt{2}}}}$, $0\le p\le m_0$, for any f\/ixed $m_0\in\mathbb{N}$. From the very def\/inition of $||\cdot ||_{\rm op}$, one infers that $||\del {\hat{a}}(m_0) ||_{\rm op}={{1}\over{{\sqrt{2}} }}$ (use for instance~\eqref{smallnorm}). Therefore, one obtains
$||[D,\pi({\hat{a}}(m_0))] ||_{\rm op}=1$ for any $m_0\in\mathbb{N}$. This proves~$iv)$.
\end{proof}

A further use of Lemma~\ref{lemme2} combined with the def\/inition of the spectral distance then permits one to determine the actual expression of the distance between two pure states $\omega_m$ and $\omega_n$.
\begin{theorem}\label{theo1}
The spectral distance between any two pure states $\omega_m$ and $\omega_n$ is
\begin{equation}
d(\omega_m,\omega_n)={\sqrt{{{\theta}\over{2}}}}\sum_{k=n+1}^m{{1}\over{{\sqrt{k}}}},\qquad \forall \, m,n\in{\mathbb{N}},\qquad n<m . \label{distancemn}
\end{equation}
It verifies the ``triangular equality''
\begin{equation}
d(\omega_m,\omega_n)=d(\omega_m,\omega_p)+d(\omega_p,\omega_n),\qquad  \forall \, m,n,p\in\mathbb{N},\qquad m\le p\le n. \label{triangular}
\end{equation}
\end{theorem}

\begin{proof}
As a consequence of Proposition \ref{relationalpha}, one can write immediately
\begin{gather*}
\alpha_{n+1,n}={\sqrt{{{n+1}\over{\theta}}}}(a_{n+1,n+1}-a_{n,n})
={\sqrt{{{n+1}\over{\theta}}}}(\omega_{n+1}(a)-\omega_{n}(a)),\qquad \forall \, n\in{\mathbb{N}} .
\end{gather*}
Then, the use of property $ii)$ of Lemma \ref{lemme2} implies that, for any $a\in{\cal{B}}_1$, one has $|\omega_{n+1}(a)-\omega_{n}(a))|\le{\sqrt{{{\theta}\over{2}}}}{{1}\over{{\sqrt{n+1}}}}$, $\forall \, n\in{\mathbb{N}}$ so that $d(\omega_{n+1},\omega_n)\le{\sqrt{{{\theta}\over{2}}}}{{1}\over{{\sqrt{n+1}}}}$, $\forall \, n\in{\mathbb{N}}$. This bound is obviously saturated by the radial element $a(n)={\sqrt{{{\theta}\over{2}}}}{{1}\over{{\sqrt{n+1}} }}f_{nn}\in{\cal{B}}_1$ as it can be immediately realized by direct calculation and using the fact that the coef\/f\/icients $\alpha_{pq}$ in the matrix base expansion of $\del a(n)$ satisfy $| \alpha_{pq}|\le\frac{1}{\sqrt{2}}$. Therefore,
\begin{equation}
d(\omega_{n+1},\omega_n)={\sqrt{{{\theta}\over{2}}}}{{1}\over{{\sqrt{n+1}}}},\qquad \forall \, n\in{\mathbb{N}}. \label{dist-interm}
\end{equation}
Now, from the general triangular inequality obeyed by the spectral distance, one must have:
\begin{gather*}
d(\omega_m,\omega_n)\le\sum_{k=n}^{m-1} d(\omega_k,\omega_{k+1}),\qquad \forall\, m,n\in\mathbb{N},
\end{gather*}
assuming for convenience $n<m$ (similar consideration obviously holds for the case $m<n$). The upper bound is indeed saturated by any element ${\hat{a}}(m_0)$, $m_0\ge n$ which belongs to ${\cal{B}}_1$. Consider $|\omega_{m}({\hat{a}}(m_0))-\omega_{n}({\hat{a}}(m_0)))|$, $n<m\le m_0$. A direct calculation yields
\begin{gather*}
|\omega_{m}({\hat{a}}(m_0))-\omega_{n}({\hat{a}}(m_0)))|={\sqrt{{{\theta}\over{2}}}}
\left|\sum_{k=m}^{m_0}{{1}\over{{\sqrt{k+1}}}}-\sum_{k=n}^{m_0}{{1}\over{{\sqrt{k+1}}}}\right|\\
\phantom{|\omega_{m}({\hat{a}}(m_0))-\omega_{n}({\hat{a}}(m_0)))|}{} =
{\sqrt{{{\theta}\over{2}}}}\sum_{k=n}^{m-1}{{1}\over{{\sqrt{k+1}}}}
=\sum_{k=n}^{m-1}d(\omega_k,\omega_{k+1}),
\end{gather*}
where \eqref{dist-interm} has been used to obtain the rightmost equality. Therefore, $d(\omega_{m},\omega_n)$ satisf\/ies \eqref{triangular}, i.e.\ the expected triangular inequality becomes an equality. The relation~\eqref{distancemn} follows immediately from~\eqref{triangular} by using~\eqref{dist-interm}.  This terminates the proof of the theorem.
 \end{proof}

\section{Discussion}\label{disc}

\subsection{Consequences of the distance formula}
We f\/irst note that another non compact spectral triple based on ${\cal{A}}^\prime=(\dlde,\star)$ has been proposed and discussed in \cite{GAYRAL2004}. We do not expect that a similar analysis of the spectral distance based on this other non compact spectral triple would noticeably alter the observations reported in this paper. Note that a part of our analysis obviously extends to $\dlde$ whenever only square integrable functions are involved.

We have shown in Theorem \ref{theo1}, equation~\eqref{triangular} that $d(\omega_m,\omega_n)=d(\omega_m,\omega_p)+d(\omega_p,\omega_n)$, for any $m,n,p\in\mathbb{N}$, $m\le p\le n$, i.e.\ the triangular inequality becomes a triangular equality. This, together with a possible relation between the family of radial functions ${\hat{a}}(m_0)$ def\/ined in $iv)$ of Lemma~\ref{lemme2} and a deformation of the positive real line of the Moyal plane has been investigated in~\cite{4-paper1}.

There are states at inf\/inite distance \cite{4-paper1}. To see that, the starting point is to use the radial element ${\hat{a}}(m_0)$ introduced above to determine a lower bound on the spectral distance between pure states given by Proposition \ref{purestates} and then show that the bound becomes inf\/inite when specif\/ic states are involved. To prepare the ensuing discussion, recall that for any unit vectors of $L^2(\mathbb{R}^2)$, $\psi=\sum_{m\in\mathbb{N}}\psi_mf_{m0}$ and $\psi^\prime=\sum_{m\in\mathbb{N}}\psi^\prime_mf_{m0}$, one infers from Proposition \ref{purestates} that
\begin{gather*}
\omega_{\psi^\prime}(a)-\omega_{\psi}(a)=2\pi\theta\sum_{m,n\in\mathbb{N}}a_{mn}(\psi^{\prime *}_m\psi^\prime_n-\psi^{ *}_m\psi_n),
\end{gather*}
for any $a\in{\cal{A}}$. Since ${\hat{a}}(m_0)$ belongs to ${\cal{B}}_1$ for any $m_0\in\mathbb{N}$, in view of $iv)$ of Lemma~\ref{lemme2}, it follows from the def\/inition~\ref{spectraldist} that the following inequality holds immediately
\begin{equation}
B(m_0;\psi,\psi^\prime):=|\omega_{\psi^\prime}({\hat{a}}(m_0))-\omega_{\psi}({\hat{a}}(m_0)) |\le d(\omega_\psi,\omega_\psi^\prime),\qquad \forall\, m_0\in\mathbb{N}. \label{bound-gene}
\end{equation}
By using the explicit expression of ${\hat{a}}(m_0)$, one obtains easily
\begin{equation}
B(m_0;\psi,\psi^\prime)=2\pi\theta{\sqrt{{{\theta}\over{2}}}}\left|\sum_{m=0}^{m_0}\sum_{k=m}^{m_0}{{1}\over{{\sqrt{k+1}} }}\big(|\psi^\prime_m |^2 - |\psi_m |^2 \big)\right|,\qquad \forall\, m_0\in\mathbb{N}. \label{bound-hat}
\end{equation}
It is convenient here to introduce some notations.

\begin{definition}\label{psiss}
Let $\psi(s)$ be a family of unit vectors of $L^2(\mathbb{R}^2)$ which is def\/ined by $\psi(s):={{1}\over{{{\sqrt{2\pi\theta}}}}} \sum_{m\in\mathbb{N}}{\sqrt{{{1}\over{\zeta(s)(m+1)^s}}}}f_{m0}$ for any $s\in\mathbb{R}$, $s>1$, where $\zeta(s)$ is the Riemann zeta function. The corresponding family of pure states are denoted by $\omega_{\psi(s)}$, for any $s\in\mathbb{R}$, $s>1$, with $\omega_{\psi(s)}$ as in Proposition~\ref{purestates}.
\end{definition}

The following property holds.
\begin{proposition}\label{state-infinite}
$d(\omega_n,\omega_{\psi(s)})=+\infty$, $\forall \,s\in\,]1,{{3}\over{2}}]$, $\forall\, n\in\mathbb{N}$.
\end{proposition}

\begin{proof}
First pick $\psi={{1}\over{{\sqrt{2\pi\theta}} }}f_{00}:=\psi_0$. Assume that $\psi^\prime=\psi(s)$. From \eqref{bound-hat}, one obtains
\begin{gather*}
B(m_0;\psi_0,\psi(s))={\sqrt{{{\theta}\over{2}}}}\left|\sum_{m=0}^{m_0}\sum_{k=m}^{m_0}{{1}\over{{\sqrt{k+1}} }}{{1}\over{\zeta(s)(m+1)^s }}-\sum_{k=0}^{m_0}  {{1}\over{{\sqrt{k+1}} }} \right|.
\end{gather*}
By splitting the sum over $k$ involved in the f\/irst term (namely, ``$\sum_{k=m}^{m_0}=\sum_{k=0}^{m_0}-\sum_{k=0}^{m}$''), this latter expression can be written as
\begin{gather}
B(m_0;\psi_0,\psi(s))={\sqrt{{{\theta}\over{2}}}}\,\Bigg|\left(1-{{1}\over{\zeta(s)}}
\sum_{m=0}^{m_0}{{1}\over{(m+1)^s}}\right)\left(\sum_{k=0}^{m_0}{{1}\over{{\sqrt{k+1}} }}\right)\nonumber\\
\phantom{B(m_0;\psi_0,\psi(s))=}{}
+{{1}\over{\zeta(s)}}\sum_{m=0}^{m_0}\sum_{k=0}^{m}{{1}\over{(m+1)^s{\sqrt{k+1}} }} \Bigg|.
\label{finalbound}
\end{gather}
From \eqref{finalbound}, one observes that $B(m_0;\psi_0,\psi(s))$ has the form $B(m_0;\psi_0,\psi(s))={\sqrt{{{\theta}\over{2}}}}|A_1(m_0)+A_2(m_0) |$ where $A_1(m_0)$ and $A_2(m_0)$ are positive terms. By further observing that
\begin{gather*}
A_2(m_0):={{1}\over{\zeta(s)}}\sum_{m=0}^{m_0}\sum_{k=0}^{m}{{1}\over{(m+1)^s{\sqrt{k+1}} }}\\
\phantom{A_2(m_0)}{} \, ={{1}\over{\zeta(s)}}\sum_{m=0}^{m_0}\sum_{k=1}^{m+1}{{1}\over{(m+1)^s{\sqrt{k}} }}\ge
{{2}\over{\zeta(s)}}\sum_{m=0}^{m_0}({{{\sqrt{m+1}}-1) }\over{(m+1)^s}},
\end{gather*}
where we used $\sum_{k=1}^{m+1}{{1}\over{{\sqrt{ k}} }}\ge 2({\sqrt{m+2}}-1)$ (see equation~\eqref{mino-1} of the appendix). $A_2(m_0)$ is bounded below by a quantity which is divergent when $m_0$ goes to $+\infty$ whenever when $s\le{{3}\over{2}}$ (owing to the property of the $\zeta$ function), one therefore concludes that $\lim\limits_{m_0\to+\infty}A_2(m_0)=+\infty$. This implies $\lim\limits_{m_0\to+\infty}B(m_0;\psi_0,\psi(s))=+\infty$. Finally, this, combined to the inequality~\eqref{bound-gene}, gives rise to $d(\omega_0,\omega_{\psi(s)})=+\infty$, $\forall\, s\in\,]1,{{3}\over{2}}]$.

Using this result, together with the triangular inequality $d(\omega_0,\omega_{\psi(s)})\!\le\! d(\omega_0,\omega_n)+d(\omega_n,\omega_{\psi(s)})$, for any $n\in\mathbb{N}$, and the fact that
$d(\omega_0,\omega_n)$ is f\/inite, in view of \eqref{distancemn}, yields $d(\omega_n,\omega_{\psi(s)})=+\infty$, $\forall\, s\in\,]1,{{3}\over{2}}]$, for any $n\in\mathbb{N}$. This terminates the proof.
\end{proof}
It turns out that the distance between states among the $\omega_{\psi(s)}$'s is inf\/inite. Indeed, the following proposition holds.
\begin{proposition}\label{infinite-points}
$d(\omega_{\psi(s_1)},\omega_{\psi(s_2)})=+\infty$, $\forall\, s_1,s_2\in \,]1,{{5}\over{4}}[   \cup  ]{{5}\over{4}},{{3}\over{2}}]$, $s_1\ne s_2$.
\end{proposition}

\begin{proof}
The proof is given in the appendix. It is based on the use of the mean value theorem to obtain a suitable estimate of the factor $(|\psi^\prime_m |^2 - |\psi_m |^2 )$ occurring in~\eqref{bound-hat}, where $\psi_m$ and $\psi^\prime_m$ are related to $\psi(s_1)$ and $\psi(s_2)$.
\end{proof}

Proposition \ref{state-infinite} shows that the topology induced by the spectral distance $d$ on the space of states of ${\bar{{\cal{A}}}}$ is not the weak $*$ topology. Having weak $*$ topology as topology of the spectral distance on the space of states is a basic condition to have compact quantum metric spaces as def\/ined in~\cite{Rieffel5}. Therefore, Proposition \ref{state-infinite} signals that the spectral triple proposed in~\cite{GAYRAL2004} is not a compact quantum metric space in the sense of~\cite{Rieffel5}. Modif\/ications of this spectral triple reinstauring the property of compact quantum metric space have been considered in~\cite{4-paper1}.

\begin{remark}\label{finitestate}
Notice that any pure state generated by a unit vector of $L^2(\mathbb{R}^2)$ built from a f\/inite linear combination of the $f_{m0}$'s stays at f\/inite distance of any pure state $\omega_n$. Indeed, let ${\cal{I}}$ be a~f\/inite subset of $\mathbb{N}$ and let $\Lambda=\sum_{m\in {\cal{I}}\subset\mathbb{N}}\lambda_mf_{m0}$ denotes a unit vector of $L^2(\mathbb{R}^2)$. Then $d(\omega_n,\omega_\Lambda)$ is f\/inite, for any $n\in\mathbb{N}$. This stems from
\begin{gather}
|\omega_\Lambda(a)-\omega_n(a)|
=\Bigg|2\pi\theta\Bigg(\sum_{p,q\in{\cal{I}}}a_{pq}\lambda_p^\star\lambda_q\Bigg)-a_{nn}\Bigg|\nonumber\\
\phantom{|\omega_\Lambda(a)-\omega_n(a)|}{} \le2\pi\theta\Bigg(\sum_{p,q\in{\cal{I}}}|a_{pq}||\lambda_p^\star\lambda_q|\Bigg)+|a_{nn}|\le
\sum_{p,q\in{\cal{I}}}|a_{pq}|+2\pi\theta|a_{nn}|, \label{distancefinite-quant}
\end{gather}
which holds for any $n\in\mathbb{N}$, and any $a\in{\cal{A}}$. In view of $iib)$ of Proposition~\ref{relationalpha}, \eqref{poincarelemma}, the $a_{mn}$'s are expressible as f\/inite sums of $\alpha_{mn}$ and $\beta_{mn}$. Furthermore, in view of $ii)$ of Lemma \ref{lemme2}, $|\alpha_{mn}|\le{{1}\over{{\sqrt{2\pi\theta}} }}$ and $|\beta_{mn}|\le{{1}\over{{\sqrt{2\pi\theta}} }}$. Therefore, the right hand side of \eqref{distancefinite-quant} is bounded.
\end{remark}

\begin{remark}
For any states $\omega_1$ and $\omega_2$, we denote by the symbol $\approx$ the equivalence relation $\omega_1\approx\omega_2\iff d(\omega_1,\omega_2)<+\infty$. $[\omega]$ denotes the equivalence class of $\omega$.
Propositions \ref{state-infinite} and~\ref{infinite-points}, Remark~\ref{finitestate}
 permit one to identify several equivalence classes.
Theorem~\ref{theo1} implies that $[\omega_n]=[\omega_0]$, $\forall \, n\in\mathbb{N}$. From Remark~\ref{finitestate}, we obtains $[\omega_\Lambda]=[\omega_0]$, where $\Lambda$ has been def\/ined in Remark~\ref{finitestate}. From Propositions~\ref{state-infinite} and~\ref{infinite-points}, $[\omega_{\psi(s_1)}]\neq[\omega_0]$, $\forall \, s_1\in\,]1,{{3}\over{2}}]$, and $[\omega_{\psi(s_1)}]\neq[\omega_{\psi(s_2)}]$, $\forall \, s_1,s_2\in\,]1,{{5}\over{4}}[   \cup  ]{{5}\over{4}},{{3}\over{2}}]$, $s_1\ne s_2$. Therefore, one has a uncountable inf\/inite family of equivalence classes. The existence of several distinct equivalent classes implies that there is no state that is at f\/inite distance to all other states: For any (pure or non pure) state, there is at least another state which is at inf\/inite distance.
\end{remark}

\subsection{Partial extension to the noncommutative torus}\label{disttorus}

It turns out that some part of the analysis presented in Section~\ref{spectrdist} and in particular the technical lemma given in Subsection~\ref{distmoy} which underlies the construction of the spectral distance formula in the Moyal case, can be actually adapted to the algebra of the noncommutative torus. This gives rise immediately to the construction of an explicit spectral distance formula between the tracial state and some vector states. We f\/irst collect the properties of the noncommutative torus that will be needed in the sequel. For more details, see \cite{ConnesTore,RieffelTore} and  \cite{CONNES,CM,LANDI,GRACIAVAR}. We assume $\theta\in{\mathbb{R}}/{\mathbb{Q}}$, $0\le\theta<1$.

\begin{definition}
${\mathfrak{A}}^2_\theta$ is the universal (unital) $C^*$-algebra generated by the two unitaries $u_1$, $u_2$ satisfying $u_1u_2=e^{i2\pi\theta}u_2u_1$. The algebra of the noncommutative torus ${\mathbb{T}}_\theta^2$ is the dense (unital) pre-$C^*$-subalgebra of ${\mathfrak{A}}^2_\theta$ def\/ined by:
\begin{gather*}
{\mathbb{T}}_\theta^2=\Bigg\{a=\sum_{i,j\in{\mathbb{Z}}}a_{ij}u_1^iu_2^j\ /\ \sup_{i,j\in{\mathbb{Z}}}\big(1+i^2+j^2\big)^k|a_{ij} |^2<\infty,\ \forall \, k\in\mathbb{N}\Bigg\} .
\end{gather*}
\end{definition}

The elements of ${\mathbb{T}}_\theta^2$ can be conveniently expressed as expansions in terms of the Weyl generators def\/ined by $U^M\equiv e^{-i\pi m_1\theta m_2}u_1^{m_1}u_2^{m_2}$, $\forall\, M=(m_1,m_2)\in{\mathbb{Z}}^2$. For any $a\in{\mathbb{T}}_\theta^2$, one can write $a=\sum_{m\in{\mathbb{Z}}^2}a_MU^M$. From now on, elements of $\mathbb{Z}^2$ (resp. $\mathbb{Z}$)  will be denoted by capital (resp.\ small) Latin letters  $M,N,\dots$ (resp.\ $m,n,\dots$) . Let $\delta_1$ and $\delta_2$ be the canonical derivations def\/ined by $\delta_a(u_b)=i2\pi u_a\delta_{ab}$, $\forall\, a,b\in\{1,2\}$. They verify $\delta_b(a^*)=(\delta_b(a))^*$, $\forall\, b=1,2$.

\begin{proposition}\label{Weylbasis}
One has for any $M,N\in{\mathbb{Z}}^2$, $(U^M)^*=U^{-M}$, $U^MU^N=\sigma(M,N)U^{M+N}$ where the commutation factor $\sigma:{\mathbb{Z}}^2\times{\mathbb{Z}}^2\to{\mathbb{C}}$ satisfies
\begin{gather*}
\sigma(M+N,P)=\sigma(M,P)\sigma(N,P),\\ \sigma(M,N+P)=\sigma(M,N)\sigma(M,P),\qquad \forall \, M,N,P\in{\mathbb{Z}}^2,
\\
\sigma(M,\pm M)=1,\qquad \forall\, M\in{\mathbb{Z}}^2,\qquad \delta_a\big(U^M\big)=i2\pi m_aU^M,\qquad  \forall \, a=1,2,\qquad  \forall \, M\in{\mathbb{Z}}^2.
\end{gather*}
\end{proposition}

\begin{proof}
The above properties can be proven by standard calculations.
\end{proof}

Let $\tau$ denotes the tracial state \cite{CONNES, GRACIAVAR}. For any $a=\sum_{M\in{\mathbb{Z}}^2}a_MU^M\in{\mathbb{T}}_\theta^2$, $\tau:{\mathbb{T}}_\theta^2\to\mathbb{C}$, $\tau(a)=a_{0,0}$. Let ${\cal{H}}_\tau$ denotes the GNS Hilbert space stemming from the completion of ${\mathbb{T}}_\theta^2$ in the Hilbert norm induced by the scalar product $\langle a,b\rangle \equiv\tau(a^* b)$. One has $\tau(\delta_b(a))=0$, $\forall \, b=1,2$.

The (compact) even real spectral triple we will consider is built from
\begin{equation}
(\mathbb{T}_\theta^2,{\cal{H}}, D; J, \Gamma), \label{tripletore}
\end{equation}
whose main properties that will be needed as collected below. For more details, see e.g.~\cite{CONNES,GRACIAVAR} and references therein. Note that the grading operator $\Gamma$ is $\Gamma={\rm diag}\, (1,-1)$. The reality operator $J$ will not enter the ensuing discussion.

The Hilbert space is ${\cal{H}}={\cal{H}}_\tau\otimes\mathbb{C}^2$. It is the Hilbert space of square integrable sections of the trivial spinor bundle over the classical torus. Notice that $\DOM(\delta_b)=\mathbb{T}_\theta^2$ and $\delta_b^\dag=-\delta_b$, $\forall\, b=1,2$, in view of $\langle \delta_b(a),c\rangle =\tau((\delta_b(a)^*c)=\tau(\delta_b(a^*)c)=-\tau(a^*\delta_b(c))=-\langle a,\delta_b(c) \rangle $ for any $b=1,2$ and $\delta_b(a^*)=(\delta_b(a))^*$. We now def\/ine $\delta=\delta_1+i\delta_2$ and ${\bar{\delta}}=\delta_1-i\delta_2$.

In \eqref{tripletore}, $D$ is the unbounded self-adjoint Dirac operator $D=-i\sum_{b=1}^2\delta_b\otimes\sigma^b$, densely def\/ined on $\DOM(D)=(\mathbb{T}_\theta^2\otimes\mathbb{C}^2)\subset{\cal{H}}$. The matrices $\sigma^b$, $b=1,2$, have been def\/ined in Section~\ref{spectrdist}. $D$ can be written as
\begin{gather*}
D=-i\begin{pmatrix} 0&\delta \\ {\bar{\delta}}& 0 \end{pmatrix}.
\end{gather*}
The faithful representation $\pi:\mathbb{T}_\theta^2\to{\cal{B}}({\cal{H}})$ is def\/ined by $\pi(a)=L(a)\otimes\bbbone_2$, $\pi(a)\psi=(a\psi_1,a\psi_2)$ for any
$\psi=(\psi_1,\psi_2)\in{\cal{H}}$, and any $a\in\mathbb{T}_\theta^2$. $L(a)$ is the left multiplication operator by any $a\in\mathbb{T}_\theta^2$. $\pi(a)$ and $[D,\pi(a)]$ are bounded operators on ${\cal{H}}$ for any $\mathbb{T}_\theta^2$. One has{\samepage
\begin{equation}
[D,\pi(a)]\psi=-i\big(L(\delta_b(a))\otimes\sigma^b\big)\psi=-i\begin{pmatrix} L(\delta(a))&0 \\ 0& L({\bar{\delta}}(a)) \end{pmatrix}\begin{pmatrix} \psi_2 \\ \psi_1 \end{pmatrix}.
\label{secondeboule}
\end{equation}
The $C^*$-completion of $\mathbb{T}_\theta^2$ is (isomorphic to) ${\mathfrak{A}}^2_\theta$. }

The def\/inition of the spectral distance between states of the torus can be immediately obtained from Def\/inition~\ref{spectraldist}. A classif\/ication of the pure states of the noncommutative torus in the irrational case is lacking. It is however possible to obtain some information on the distance between any of the particular vector states as def\/ined below and the tracial state.

In the present case, the set ${\cal{B}}_1$, Def\/inition~\ref{unit-ball} can be conveniently characterized in a way somewhat similar to the case of Moyal planes by the following lemma.
\begin{lemma}\label{lemme4}
Set $\delta(a)=\sum_{N\in\mathbb{Z}^2}\alpha_NU^N$. One has $\alpha_N=i2\pi(n_1+in_2)a_N$, $\forall\, N=(n_1,n_2)\in\mathbb{Z}^2$.

$i)$ For any $a\in{\cal{B}}_1$, $||[D,\pi(a)] ||_{\rm op}\le1$ implies $|\alpha_N |\le1$, $\forall\, N\in\mathbb{Z}^2$. Similar results hold for~${\bar{\delta}}(a)$.

$ii)$ The elements ${\hat{a}}^M\equiv{{U^M}\over{2\pi(m_1+im_2)}}$ verify $||[D,\pi({\hat{a}}^M)] ||_{\rm op}=1$, $\forall\, M=(m_1,m_2)\in\mathbb{Z}^2$, $M\ne(0,0)$.
\end{lemma}

\begin{proof}
The relation involving $\alpha_N$ is obtained from an elementary calculation. Then \mbox{$\forall\, a\in {\cal{B}}_1$}, $||[D,\pi(a)] ||_{\rm op}\le1$ is equivalent to $||\delta(a) ||_{\rm op}\le1$ and
$||{\bar{\delta}}(a) ||_{\rm op}\le1$ in view of \eqref{secondeboule}. For any~\mbox{$a\in{\mathfrak{A}}^2_\theta$} and any unit $\psi=\sum_{N\in\mathbb{Z}^2}\psi_NU^N\in{\cal{H}}_\tau$, one has $||\delta(a)\psi ||^2=\sum_{N\in\mathbb{Z}^2}|\sum_{P\in\mathbb{Z}^2}\alpha_P\psi_{N-P}\sigma(P,N) |^2$. Then $||\delta(a) ||_{\rm op}\le1$ implies $|\sum_{P\in\mathbb{Z}^2}\alpha_P\psi_{N-P}\sigma(P,N) |\le1$, for any $N\in\mathbb{Z}^2$ and any unit $\psi\in{\cal{H}}_\tau$. By a straightforward adaptation of the proof carried out for $ii)$ of Lemma~\ref{lemme2}, this implies $|\alpha_M|\le1$, $\forall\, M\in\mathbb{Z}^2$. This proves $i)$. Finally, $ii)$ stems simply from an elementary calculation.
\end{proof}

We now determine the distance between the states mentioned above.
\begin{proposition}
Let the family of unit vectors $\Phi_M=\big({{1+U^M}\over{{\sqrt{2}}}} ,0 \big)\in{\cal{H}}$, $\forall\, M\in\mathbb{Z}^2$, $M\ne(0,0)$ generating the family of vector states of $\mathbb{T}_\theta^2$
\begin{gather*}
\omega_{\Phi_M}:\ \mathbb{T}_\theta^2\to\mathbb{C},\qquad \omega_{\Phi_M}(a)\equiv( \Phi_M,\pi(a)\Phi_M)_{{\cal{H}}}=\tfrac{1}{2}\langle (1+U^M),(a+aU^M) \rangle.
\end{gather*}
The spectral distance between any state $\omega_{\Phi_M}$ and the tracial state is
\begin{equation}
d(\omega_{\Phi_M},\tau)={{1}\over{2\pi|m_1+im_2|}},\qquad \forall\, M=(m_1,m_2)\in\mathbb{Z}^2,\qquad M\ne(0,0). \label{distfixe}
\end{equation}
\end{proposition}

\begin{proof}
Set $a=\sum_{N\in\mathbb{Z}^2}a_NU^N$. By using properties of the Weyl generators, an elementary calculation yields $\omega_{\Phi_M}(a)=\tau(a)+{{1}\over{2}}(a_M+a_{-M})$. This, combined with Lemma \ref{lemme4} leads immediately to $d(\omega_{\Phi_M},\tau)\le{{1}\over{2\pi|m_1+im_2 |}}$.
Now, this upper bound is obviously saturated by the element ${\hat{a}}^M$ of Lemma~\ref{lemme4}. The result~\eqref{distfixe} follows.
\end{proof}

\section{Conclusion}\label{section4}

We have studied the spectral distance for two-dimensional noncommutative Moyal planes in the framework of the non compact spectral triple proposed in  \cite{GAYRAL2004} as a possible noncommutative analog of non compact Riemannian spin manifold. The spectral distance can be explicitly determined between any elements of a particular class of pure states. An explicit formula for the corresponding distance has been given. There exists a family of pure states that are at inf\/inite distance from each other as well as at inf\/inite distance from some classes of pure states, including those for which the constructed spectral distance formula is valid. The fact that the spectral distance can be used here to def\/ine an equivalence relation among (pure) states is brief\/ly discussed. Note that for any state, there is at least another state which is at inf\/inite distance from it. The existence of some pure states that are at inf\/inite distance implies that the topology induced by the spectral distance on the space of states of the algebra involved in the triple is not the weak $*$ topology. Therefore, the noncompact spectral triple proposed in~\cite{GAYRAL2004}  does not give rise to a compact quantum metric spaces as def\/ined in~\cite{Rieffel5}. The case of the noncommutative torus is brief\/ly considered and a formula for the spectral distance between the tracial state and some vector states is obtained.

\appendix
\section{Proof of Proposition \ref{infinite-points} }
In this appendix, we give a detailed proof of Proposition \ref{infinite-points}.
\begin{proof}
Let us assume $s_1<s_2$, with $s_i\in\,]1,{{3}\over{2}}]$, $i=1,2$ according to Section~\ref{disc}. The analysis for the case $s_2<s_1$ is completely similar.
It is convenient here to def\/ine additional notations. The other notations have been already def\/ined in Section~\ref{disc}.
\begin{definition}\label{quantities}
For $1<s_1<s_2\le {{3}\over{2}}$, $0\le m\le m_0$, $m,m_0\in\mathbb{N}$, we set:
\begin{gather*}
u(m,m_0):=\sum_{k=m}^{m_0}{{1}\over{{\sqrt{k+1}}}},\qquad G_{s_1,s_2}(m):=\left({{1}\over{\zeta(s_1)(m+1)^{s_1}}} -{{1}\over{\zeta(s_2)(m+1)^{s_2}}} \right).
\end{gather*}
\end{definition}
We now prove some estimates that will be needed in the ensuing analysis.
\begin{proposition}\label{estimate}
The following estimates hold:
\begin{gather}
\big({\sqrt{m_0+2}}-{\sqrt{m+1}}\big)\le{{1}\over{2}} \sum_{k=m}^{m_0}{{1}\over{{\sqrt{k+1}} }}\nonumber\\
\hphantom{\big({\sqrt{m_0+2}}-{\sqrt{m+1}}\big)}{}
\le \big({\sqrt{m_0+1}}-{\sqrt{m}}\big),\qquad \forall \, m,m_0\in\mathbb{N},\qquad m<m_0,\label{estim-1}
\\
{{1}\over{(A+1)^{s-1}}}-{{1}\over{(m_0+2)^{s-1}}}\le(s-1)\sum_{m=A}^{m_0}{{1}\over{(m+1)^s}}\nonumber\\
\hphantom{{{1}\over{(A+1)^{s-1}}}-{{1}\over{(m_0+2)^{s-1}}}}{}
\le{{1}\over{(A)^{s_1}}}-{{1}\over{(m_0+1)^{s-1}}},\qquad \forall \, s\in \big]1,\tfrac{3}{2}\big],\label{estim-2}
\end{gather}
where $A\in\mathbb{N}$, $A<m_0$.
\end{proposition}
\begin{proof}
The application of the mean value theorem to the function $f(x)=x^\alpha$, $\alpha\in\mathbb{R}$, $\forall\, x\in[k,k+1]$, $\forall \, k\in\mathbb{N}$ gives rise to:
\begin{gather}
\alpha k^{\alpha-1}\ge(k+1)^{\alpha}-k^\alpha \ge \alpha(k+1)^{\alpha-1},\qquad \forall\, \alpha\in\, ]0,1],\qquad k\ne0\label{mino-1},
\\
\alpha k^{\alpha-1}\le(k+1)^{\alpha}-k^\alpha \le \alpha(k+1)^{\alpha-1},\qquad \forall \, \alpha<0,\qquad k\ne0\label{mino-2}.
\end{gather}
Set respectively $\alpha={{1}\over{2}}$ and $\alpha=1-s$, $1<s\le {{3}\over{2}}$ in \eqref{mino-1} and \eqref{mino-2}. Proposition \ref{estimate} follows.
\end{proof}

Let $u_n$ and $v_n$ be two sequences. We def\/ine some notations. We set: $v_n=O(u_n)$ if $\exists \, C>0$,  / $\forall\,  n\in\mathbb{N}$, $|v_n|<C|u_n|$, and $v_n=o(u_n)$ if $\forall\, \epsilon>0$, $\exists\, N\in\mathbb{N}$, / $\forall\, n>N$, $|v_n|<\epsilon|u_n|$. Recall that for any $\in\mathbb{N}$, $u_n\neq 0$, $v_n=O(u_n)$ means that, as $n$ approaches the inf\/inity,  $\frac{v_n}{u_n}$ is bounded while $v_n=o(u_n)$ means that $\frac{v_n}{u_n}\to 0$. We further recall that if  $v_n=O(u_n)$ and the serie $\sum_{k=0}^{n}|v_n|\to\infty$, as $n\to\infty$, then $\sum_{k=0}^{n}|u_n|\to\infty$ and $\sum_{k=0}^{n}|v_n|=O\left(\sum_{k=0}^{n}|u_n|\right)$ as $n\to \infty$. Similarly, if $v_n=O(u_n)$ and the serie $\sum_{k=0}^{n}|u_n|$ is convergent as $n\to\infty$, then $\sum_{k=0}^{n}|v_n|$ and the rests verify $\sum_{k=n}^{\infty}|v_n|=O\left(\sum_{k=n}^{\infty}|u_n|\right)$ as $n\to \infty$. A similar property holds for  $v_n=o(u_n)$.

Now, by using \eqref{bound-hat} and Def\/inition \ref{psiss}, one obtains
\begin{equation}
B(m_0;\psi(s_1),\psi(s_2))={\sqrt{{{\theta}\over{2}}}}\left|\sum_{m=0}^{m_0} u(m,m_0)G_{s_1,s_2}(m) \right|.\label{B-interm}
\end{equation}
We def\/ine $M$ with the following condition:
$G_{s_1,s_2}(m)\le0$ when $0\le m\le M$ and $G_{s_1,s_2}(m)\ge0$ otherwise ($M$ can be determined from the equation $\zeta(s_2)(m+1)^{s_2}=\zeta(s_1)(m+1)^{s_1}$).

Using $\sum_{m=0}^{\infty}G_{s_1,s_2}(m)=0$, we def\/ine (for $m_0>M$)
\begin{gather*}
B^-(m_0,s_1,s_2)=\sum_{m=0}^{M}u(m,m_0)G_{s_1,s_2}(m),
\\
B^+(m_0,s_1,s_2)=\sum_{m=M+1}^{m_0}u(m,m_0)G_{s_1,s_2}(m),
\\
\alpha=\sum_{m=M+1}^{\infty}G_{s_1,s_2}(m)=-\sum_{m=0}^{M}G_{s_1,s_2}(m)>0.
\end{gather*}
It is easy to observe that \eqref{B-interm} can be written as
\begin{equation}
B(m_0;\psi(s_1),\psi(s_2))={\sqrt{{{\theta}\over{2}}}}\left| B^-(m_0,s_1,s_2)+B^+(m_0,s_1,s_2)\right|.\label{B-interm-1}
\end{equation}
To show that the limit of $B(m_0;\psi(s_1),\psi(s_2))$ is equal to $+\infty$ as $m_0\to+\infty$, one has to compare the corresponding asymptotic behavior of the terms built from the series involved in~\eqref{B-interm-1}. We f\/irst consider the f\/irst term~$B^-$. By using the estimate~\eqref{estim-1}, one obtains
\begin{gather}
2\sum_{m=0}^{M}\big({\sqrt{m_0+1}}-{\sqrt{m}}\big)G_{s_1,s_2}(m) \nonumber\\
\qquad{}
\le B^-(m_0,s_1,s_2) \le 2\sum_{m=0}^{M}\big({\sqrt{m_0+2}}-{\sqrt{m+1}}\big)G_{s_1,s_2}(m), \label{encad-1}
\\
-2\big({\sqrt{m_0+1}}\big)\alpha + C_1\le B^-(m_0,s_1,s_2) \le  -2\big({\sqrt{m_0+2}}\big)\alpha + C_2,\nonumber
\end{gather}
where $C_1$ and $C_2$ are constant.
Then, it follows that $B^-(m_0,s_1,s_2)=-2({\sqrt{m_0+2}})\alpha+O(1)$ as $m_0\to+\infty$ because of ${\sqrt{m_0+2}}-{\sqrt{m_0+1}}=O({m_0}^{\frac{-1}{2}})$.

Consider now $B^+$. By using the estimate~\eqref{estim-1}, one can write a relation similar to~\eqref{encad-1}:
\begin{gather*}
2\sum_{m=M+1}^{m_0}\big({\sqrt{m_0+2}}-{\sqrt{m+1}}\big)G_{s_1,s_2}(m)\nonumber\\
\qquad{} \le B^+(m_0,s_1,s_2) \le 2\sum_{m=M+1}^{m_0}\big({\sqrt{m_0+1}}-{\sqrt{m}}\big)G_{s_1,s_2}(m).
\end{gather*}
 In this latter relation, the leftmost serie can be expressed as
\begin{gather*}
S_L:=2\sum_{m=M+1}^{m_0}\big({\sqrt{m_0+2}}-{\sqrt{m+1}}\big)G_{s_1,s_2}(m)
\\
\phantom{S_L}{} \, =2\left(\sum_{m=M+1}^{m_0}{\sqrt{m_0+2}}G_{s_1,s_2}(m)
-\sum_{m=M+1}^{m_0}G_{s_1-{{1}\over{2}},s_2-{{1}\over{2}}}(m)\right)
\\
\phantom{S_L}{} \,
=2(\sqrt{m_0+2})\alpha -2\left({\sqrt{m_0+2}}\sum_{m=m_0}^{\infty}G_{s_1,s_2}(m)+\sum_{m=M+1}^{m_0}G_{s_1-{{1}\over{2}},
s_2-{{1}\over{2}}}(m)\right).
\end{gather*}
As $m$ approaches the inf\/inity, one has $G_{s_1-{{1}\over{2}},s_2-{{1}\over{2}}}(m)={{1}\over{\zeta(s_1)m^{s_1-{{1}\over{2 }}} }}+O\left({{1}\over{m^{s_2-{{1}\over{2 }}} }}\right)$. The quantity $\sum_{m=M+1}^{m_0}G_{s_1-{{1}\over{2}},s_2-{{1}\over{2}}}(m)$ is therefore a diverging serie as $m_0$ approaches the inf\/inity ( $s_i-{{1}\over{2}}\le 1$, $i=1,2$) and
\begin{gather*}
\sum_{m=M+1}^{m_0}G_{s_1-{{1}\over{2}},s_2-{{1}\over{2}}}(m)
=
\sum_{m=M+1}^{m_0}{{1}\over{\zeta(s_1)m^{s_1-{{1}\over{2 }}} }}
+O\left(\sum_{m=M+1}^{m_0}{{1}\over{m^{s_2-{{1}\over{2 }}} }}\right)
\\
\hphantom{\sum_{m=M+1}^{m_0}G_{s_1-{{1}\over{2}},s_2-{{1}\over{2}}}(m)}{}
={{1}\over{\zeta(s_1)(s_1-{{3}\over{2}})}}m_0^{{{3}\over{2}}-s_1}
+O\big(m_0^{{{3}\over{2}}-s_2}\big),
\end{gather*}
where the last equality can be checked from \eqref{estim-2}. Conversely, because of $G_{s_1,s_2}(m)={{1}\over{\zeta(s_1)m^{s_1}}} +O\left({{1}\over{m^{s_2}}}\right)$ as $m\to\infty$, the quantity $\sum_{m=0}^{m_0}G_{s_1,s_2}(m)$ is a converging serie as $m_0$ approaches the inf\/inity and we can compare the rests of this series:
\begin{gather*}
\sum_{m=m_0}^{\infty}G_{s_1-{{1}\over{2}},s_2-{{1}\over{2}}}(m)
=\sum_{m=m_0}^{\infty}{{1}\over{\zeta(s_1)m^{s_1} }}
+O\left(\sum_{m_0}^{\infty}{{1}\over{m^{s_2} }}\right)
\\
\hphantom{\sum_{m=m_0}^{\infty}G_{s_1-{{1}\over{2}},s_2-{{1}\over{2}}}(m)}{}
={{1}\over{\zeta(s_1)(s_1-1)}}m_0^{1-s_1}
+O\big(m_0^{1-s_2}\big).
\end{gather*}
From this, it follows that
\begin{gather*}
S_L=2\big(\sqrt{m_0+2}\big)\alpha -\frac{2}{\zeta(s_1)}\left( {{1}\over{(s_1-1)}} +{{1}\over{(s_1-{{3}\over{2}})}}\right)m_0^{{{3}\over{2}}-s_1} +O\big(m_0^{{{3}\over{2}}-s_2}\big).
\end{gather*}
A similar analysis can be carried out for the rightmost serie. We therefore can write
\begin{gather*}
B^+(m_0,s_1,s_2)=2\big(\sqrt{m_0+2}\big)\alpha -\frac{2}{\zeta(s_1)}\left( {{1}\over{(s_1-1)}} +{{1}\over{(s_1-{{3}\over{2}})}}\right)m_0^{{{3}\over{2}}-s_1} +O\big(m_0^{{{3}\over{2}}-s_2}\big)
\\
\hphantom{B^+(m_0,s_1,s_2)}{} =2\big(\sqrt{m_0+2}\big)\alpha -\frac{2}{\zeta(s_1)}\left(\frac{(2s_1-\frac{5}{2})}{(s_1-1)(s_1-\frac{3}{2})}\right)m_0^{{{3}\over{2}}-s_1} +O\big(m_0^{{{3}\over{2}}-s_2}\big),
\\
B^-(m_0,s_1,s_2)+B^+(m_0,s_1,s_2)=-\frac{2}{\zeta(s_1)}
\left(\frac{(2s_1-\frac{5}{2})}{(s_1-1)(s_1-\frac{3}{2})}\right)m_0^{{{3}\over{2}}-s_1} +O\big(m_0^{{{3}\over{2}}-s_2}\big).
\end{gather*}
Therefore, $B(m_0;\psi(s_1),\psi(s_2))\to\infty$ if $s_1\neq\frac{5}{4}$ and
$d(\omega_{\psi(s_1)},\omega_{\psi(s_2)})=+\infty$, $\forall \, s_1,s_2\in \,]1,{{5}\over{4}}[   \cup  ]{{5}\over{4}},{{3}\over{2}}]$, $s_1\ne s_2$.
This terminates the proof.
\end{proof}

\begin{remark}
We notice that for $s_2\neq\frac{3}{2}$, it can be shown that
\begin{gather*}
B^-(m_0,s_1,s_2)+B^+(m_0,s_1,s_2)
=-\frac{2}{\zeta(s_1)}\left(\frac{(2s_1-\frac{5}{2})}{(s_1-1)(s_1-\frac{3}{2})}\right)m_0^{{{3}\over{2}}-s_1}
\\
\phantom{B^-(m_0,s_1,s_2)+B^+(m_0,s_1,s_2)=}{}
+\frac{2}{\zeta(s_2)}\left(\frac{(2s_2-\frac{5}{2})}{(s_2-1)(s_2-\frac{3}{2})}\right)m_0^{{{3}\over{2}}-s_2} +o\big(m_0^{{{3}\over{2}}-s_2}\big).
\end{gather*}
From this, one obtains that
$d(\omega_{\psi(s_1)},\omega_{\psi(s_2)})=+\infty$, $\forall\, s_1,s_2\in\, ]1,{{3}\over{2}}[$, $s_1\ne s_2$.
However, one cannot conclude for $d(\omega_{\psi(\frac{5}{4})},\omega_{\psi(\frac{3}{2})})$. Therefore, either $s=\frac{5}{4}$ or $\frac{3}{2}$ must be removed from the set $]1,{{3}\over{2}}]$ in order to have a family of states for which the distance $d(\omega_{\psi(s_1)},\omega_{\psi(s_2)})$ is known.
\end{remark}

\subsection*{Acknowledgments}

One of us (E.C.) thanks the organizers of the XVIIIth International Colloquium on Integrable Systems and Quantum Symmetries for their kind invitation which motivated the present paper. We thank the referees for their fruitful suggestions and comments that helped us to improve the initial version of this paper. Discussions and correspondence on various metric aspects of noncommutative geometry with F.~d'Andrea and P.~Martinetti are gratefully acknowledged. One of us (JCW) thanks F.~Lizzi and B.~Iochum for discussions and comments. We thank E.~Jolibois for discussions at various stage of this work.

\pdfbookmark[1]{References}{ref}
\LastPageEnding

\end{document}